\documentclass[twocolumn]{aastex63}

\usepackage{ulem}
\accepted{18-Nov-2020 to ApJL}
\shorttitle{Pulsational pair-instability}
\shortauthors{Umeda et al.}
\graphicspath{{./}{figures/}}

\begin{document}

\title{Pulsational pair-instability and the mass gap of Population III 
Black Holes: Effects of overshooting}

\correspondingauthor{Hideyuki Umeda}
\email{umeda@astron.s.u-tokyo.ac.jp}

\author[0000-0001-8338-502X]{Hideyuki Umeda}
\affiliation{Department of Astronomy, Graduate School of Science, The University of Tokyo, 7-3-1 Hongo, Bunkyo-ku, Tokyo 113-0033, Japan}

\author[0000-0002-8967-7063]{Takashi Yoshida}
\affil{Department of Astronomy, Graduate School of Science, University of Tokyo, 7-3-1 Hongo, Bunkyo-ku, Tokyo 113-0033, Japan}

\author{Chris Nagele}
\affil{Department of Astronomy, Graduate School of Science, University of Tokyo, 7-3-1 Hongo, Bunkyo-ku, Tokyo 113-0033, Japan}

\author[0000-0002-6705-6303]{Koh Takahashi}
\affiliation{Max Planck Institute for Gravitational Physics, D-14476 Potsdam, Germany}






\begin{abstract}
Since the discovery of GW190521, several proposals have been put forward 
to explain the formation of a black hole (BH) in the mass 
gap caused by (pulsational) pair-instability, $M = 65-130 M_\odot$. We calculate 
the mass ejection of Population III 
stars by the pulsational pair-instability (PPI) 
process using a stellar evolution and hydrodynamical code.
If a relatively small, but reasonable value is adopted 
for the overshooting parameter, the stars do not become
red super giants during the PPI phase. We show that in this case
most of the hydrogen envelope remains after the mass ejection by PPI.
We find that the BH mass could be at most around $110 M_\odot$ 
below the mass range of pair-instability supernovae.


\end{abstract}



\section{Introduction} \label{sec:Intro}
The discovery of GW190521 \citep{2020PhRvL.125j1102A}
raised an important question. It was suggested that the primary
back hole (BH) was in the BH mass gap, 
$M = 65 - 130 M_\odot$ \citep[e.g.,][]{2019ApJ...887...53F, 2020ApJ...888...76M}, which
is caused by the (pulsational) pair-instability (P)PI process
\citep[e.g.,][]{2017ApJ...836..244W}.
This primary mass may be explained if a BH grows after its formation.
Several such formation scenarios have been proposed. For example,
BH mergers in star
clusters \citep[e.g.][]{2019PhRvD.100d3027R, 2020MNRAS.497.1043D, 2020ApJ...902L..26F, 2020arXiv200911447L}, gas accretion onto Population (Pop) III
BHs \citep[e.g.][]{2019A&A...632L...8R, 2020ApJ...903L..21S}
and combination of BH mergers and gas accretion in disks of active galactic nuclei \citep[e.g.][]{2019PhRvL.123r1101Y,2020ApJ...898...25T}. 

On the other hand,
\cite{2020arXiv200906585F}, \cite{2020arXiv200906922K} and \cite{2020arXiv201007616T} 
discuss that the binary evolution model may explain a BH mass larger than 
$65M_\odot$ under some assumptions. Specifically, \cite{2020arXiv201007616T}
claims that a larger BH mass can be realized if a star does not become
a red super giant due to a relatively small overshooting parameter. 
It was also suggested that if a small $^{12}$C$(\alpha, \gamma)^{16}$O 
reaction rate is adopted,
the mass range for the PPI phase shrinks and can even disappear
(\citealt{2020ApJ...902L..36F}; see also \citealt{2018ApJ...863..153T}). In this case the BH mass gap
can be shrunk to explain GW190521 \citep{2020arXiv200913526B}. 

 In this $Letter$, we calculate PPI mass loss 
 using a stellar evolution and hydrodynamical code to re-investigate
 the BH mass gap for Pop III stars.
 We note that although the PPI mass loss has been discussed by several
 authors, no previous works have actually calculated the mass loss
 process using Pop III stellar evolution models. For example, in the
 work of \cite{2017ApJ...836..244W}, 
 metal free models are mimicked 
 by setting the mass loss rate of a metal poor model to be zero. 
 We show that the BH mass 
 gap depends sensitively on the overshooting parameter. The 
 lower bound of the gap can be much larger than previously thought
 if we adopt a relatively small, but reasonable, overshooting parameter.  


\section{Method} \label{sec:Method}

 The calculation method is similar to 
\cite{2016MNRAS.457..351Y}.
We first calculate stellar evolution using the HOSHI code 
\citep{2018ApJ...857..111T}
including the PPI phase. In this $Lettter$, we adopt a 49 isotope nuclear reaction network. Note that in the HOSHI code, acceleration terms are not
solved and thus hydrodynamical evolution can not be solved accurately.  
During the PPI phase, mass ejection may occur due to hydrodynamical
 effects. We solve such mass ejection using a piecewise parabolic method hydrodynamical code, e.g., \citet{1984JCoPh..54..174C},
including nuclear energy generation.
If mass loss occurs we calculate the evolution after removing 
the corresponding mass from the surface of the star. 

The models we calculate are Pop III stars with zero metallicity
and have initial masses in the range, $M = 70 - 135 M_\odot$.
The $^{12}$C$(\alpha, \gamma)^{16}$O is chosen to be 1.5 times the 
CF88 rate \citep{1988ADNDT..40..283C}
since, as described in 
\cite{2012PTEP.2012aA302U},
the solar abundance ratios are reproduced well 
if the rate is slightly larger
than the value adopted there, 1.3 times the CF88 rate.

In this $Letter$, we focus on the effect of overshooting. 
In the HOSHI code, the overshooting is taken with a diffusive
treatment. We consider
two cases for the overshooting parameter $f_{\rm OV}$ which is defined
e.g., in \citet{Yoshida19}. Following that paper, we call
the $f_{\rm OV}$=0.01 and 0.03 cases M and L models, respectively.
The M model is a similar choice to GENEC 
\citep{2012A&A...537A.146E}, and
the L model is similar to Stern \citep{2011A&A...530A.115B}.

\section{Results} \label{sec:Results}

Here we first describe typical differences between the L and M models. 
Fig. 1 shows the HR-diagrams of 100$M_\odot$
models. The L models tend to be red after H-burning, 
while the M models stay blue in the mass range studied here. The differences
between the two models are not large for lower masses ($M < 30M_\odot$)
but are signifigant for more massive stars,  
\citep[e.g.,][]{2020arXiv201007616T}.

Figure 2 shows the evolution of central temperature $T_{\rm c}$ and
central density $\rho_{\rm c}$ for the 100$M_\odot$ models
around the PPI phase. Both
models show temperature \& density oscillations during 
the central Si-burning phase.
This oscillation phase is commonly called the PPI phase. As shown below
only the last few large oscillations can cause mass ejection. 

In Figure 3, we show the internal density and radius distribution
of the 100$M_\odot$ models
as a function of enclosed mass ($M_r$) at a time just before the first mass ejection: Log $T_{\rm c}$ = 9.72 (M model) 
and Log $T_{\rm c}$ = 9.64 (L model).
We find that inner structures ($M_r < 40 M_\odot$) are similar, but
the envelope structures ($M_r > 40 M_\odot$) are quite different.

Table \ref{tab:tab1} summarizes the main results.
$M_{\rm ini}$, $M_{\rm CO}$ and $M_{\rm He}$ are
the initial, final CO-core and final He-core masses.
For most L models, the hydrogen mass
fraction changes rapidly at the edge of the He core so that the lower and
upper bounds of $M_{\rm He}$ are identical. 
The next column shows the number of major PPI oscillations which have peak temperatures Log $T_{\rm peak} > 9.59$. The restriction is made because we find mass ejection can only accompany major oscillations, and specifically those with high peak temperatures, although numerous smaller oscillations can occur as shown in Fig.2. 
Furthermore, not all of the major oscillations result in mass ejections, and we find that the number of mass ejections, which are listed in the next column, is at most 2.
We think this is one of the most interesting results in this $Letter$.
For the L models, we find only one mass ejection at most.
This is in contrast to the case of CO stars 
\citep{2016MNRAS.457..351Y}. For
Pop III stars, most shock waves produced by PPI, 
excluding the last one or two, are damped out
in the hydrogen envelope without mass ejection.
The peak temperature during a PPI oscillation accompanying mass ejection
is shown in the next column. The last two columns are the
remnant mass after the mass ejection and the energy of the ejecta, respectively.

\begin{deluxetable*}{cccccccc}
\tablenum{1}
\tablecaption{Summary of the results \label{tab:tab1}}
\tablewidth{0pt}
\tablehead{\colhead{$M_{\rm ini}$} & $M_{\rm CO}$ & $M_{\rm He}$ & \# of PPI &
Ejection \# & Log $T_{\rm peak}$ & $M_{\rm rem}$ & Ejecta Energy \\
\colhead{($M_\odot$)} & \colhead{($M_\odot$)} & \colhead{($M_\odot$)} & \colhead{} & \colhead{} & \colhead{(K)}& \colhead{($M_\odot$)} & 
\colhead{($\rm 10^{50} erg$)} 
}
\startdata
\multicolumn{8}{c}{L Models ($f_{\rm OV}$=0.03)} \\
70 & 34.2 & 38.9-48.8 & 4 & 0 &  -  & 70 & - \\
75 & 34.9 & 39.3 & 4 & 1 & 9.81 & 42.4 & 6.5 \\
80 & 37.4 & 42.2-42.9 & 3 & 1 & 9.71 & 42.4 & 0.18 \\
100 & 48.1 & 53.6 & 2 & 1 & 9.65 & 52.2 & 4.5 \\
120 & 57.9 & 64.9 & 1 & 1 & 9.66 & 60.3 & 4.7 \\
135 & 65.4 & 73.5 & 1 & 1 & 9.63 & 66.9 & 5.6 \\
\hline
\multicolumn{8}{c}{M Models ($f_{\rm OV}$=0.01)} \\
70 & 27.0 & 30.3-34.4 & 0 & 0 & - & 70 & - \\
80 & 31.8 & 35.3-39.4 & 5 & 0 & - & 80 & - \\
90 & 37.2 & 41.9-44.8 & 3 & 1 & 9.76 & 83.0 & 1.4 \\
100 & 42.7 & 47.3-52.1 & 2 & 1 & 9.73 & 91.7 & 1.5 \\
110 & 46.7 & 50.8-56.5 & 2 & 1 & 9.60 & (105.5) & 0.44 \\
    & \multicolumn{3}{l}{(interval: 0.064 yr)} 
    & 2 & 9.72 & 91.9 & 3.5 \\
115 & 47.9 & 55.1-62.7 & 1 & 1 & 9.69 & 99.0 & 5.5 \\
120 & 50.8 & 56.3-71.3 & 2 & 1 & 9.64 & (107.0)& 3.4 \\
    & \multicolumn{3}{l}{(interval: 0.82 yr)} 
    & 2 & 9.86 & 69.9 & 49 \\
125 & 55.9 & 63.9-64.9 & 2 & 1 & 9.61 & (99.5) & 6.5 \\
    & \multicolumn{3}{l}{(interval: 3.1 yr)}   
    & 2 & 9.79 & 65.1 & 52 \\
130 & 55.6 & 60.9-75.1 & 2 & 1 & 9.60 & (114.3) & 4.7 \\
    & \multicolumn{3}{l}{(interval: 3.1 yr)}    
    & 2 & 9.80 & 77.7 & 39 \\
135 & 58.3 & 65.3-72.6 & 1 & 1 & 9.64 & 108.7 & 5.6 \\
\enddata
\tablecomments{$M_{\rm ini}$, $M_{\rm CO}$ and $M_{\rm He}$ are
the initial, final CO-core and final He-core masses.
$M_{\rm CO}$ is defined as the enclosed mass of the CO rich core with
helium mass fraction $X({\rm He}) <0.01$. The range of $M_{\rm He}$ is
defined by $X({\rm H}) <0.1$ for the lower bound and
$<0.3$ for the upper bound. The next column is the number of 
PPI oscillations which have peak temperatures Log $T_{\rm peak} > 9.59$. 
The next column is the number of mass ejections. Here, 1 and
2 represent the first and second mass ejections.
For the second ejection, we also show the time interval between 
the two peaks.
The peak temperature during a mass ejecting pulse
is shown next. The last two columns are the remnant mass and energy
of the ejecta. $M_{\rm rem}$ in the parentheses is 
the stellar mass after the first pulse.}
\end{deluxetable*}

\begin{figure}[ht]
\includegraphics[width=8.5cm]{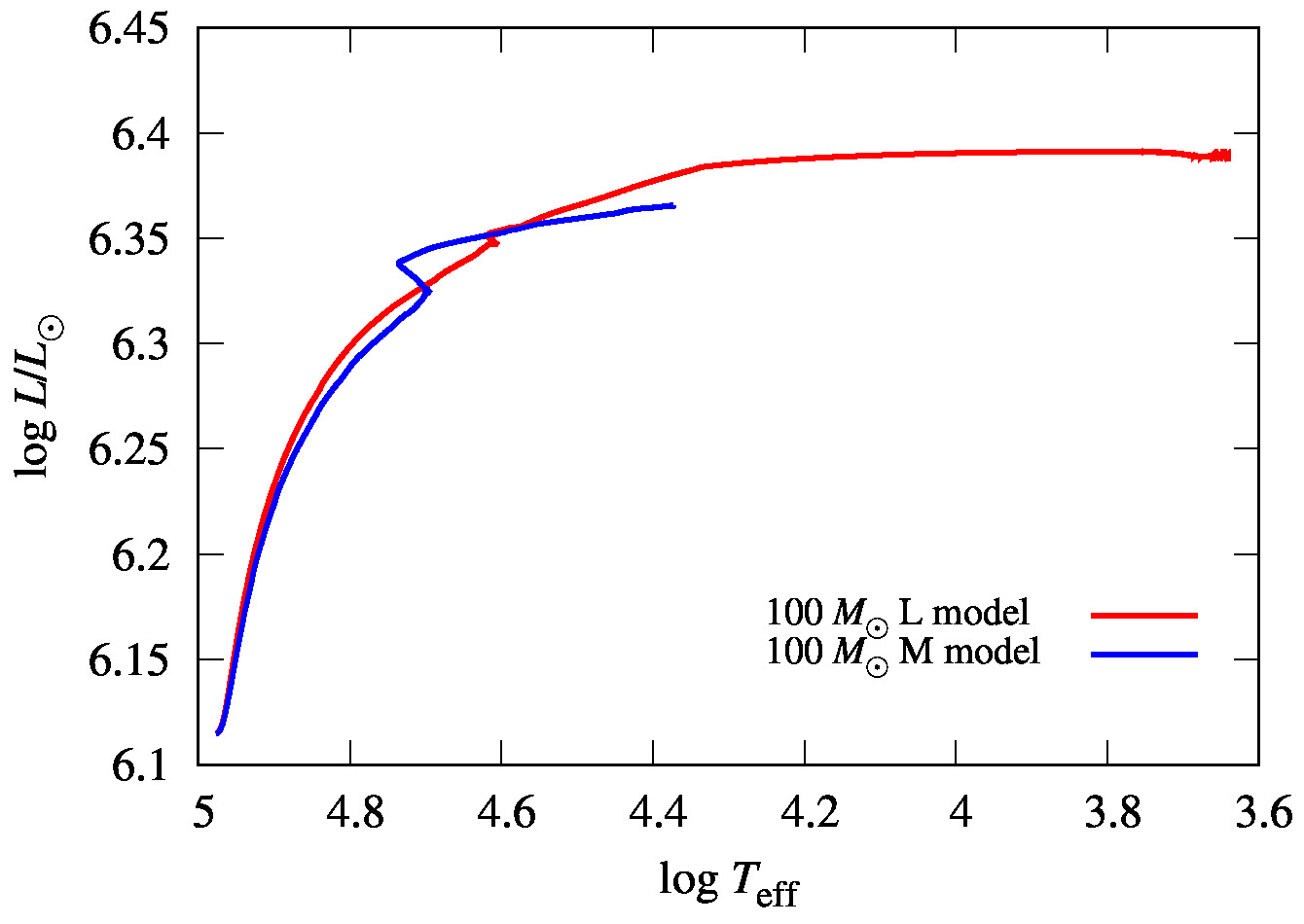}
\caption{The HR-diagram of 100$M_\odot$ models. 
The red and blue curves indicate L and M models, respectively.
\label{fig:1}}
\end{figure}

\begin{figure}[ht]
\includegraphics[width=8.5cm]{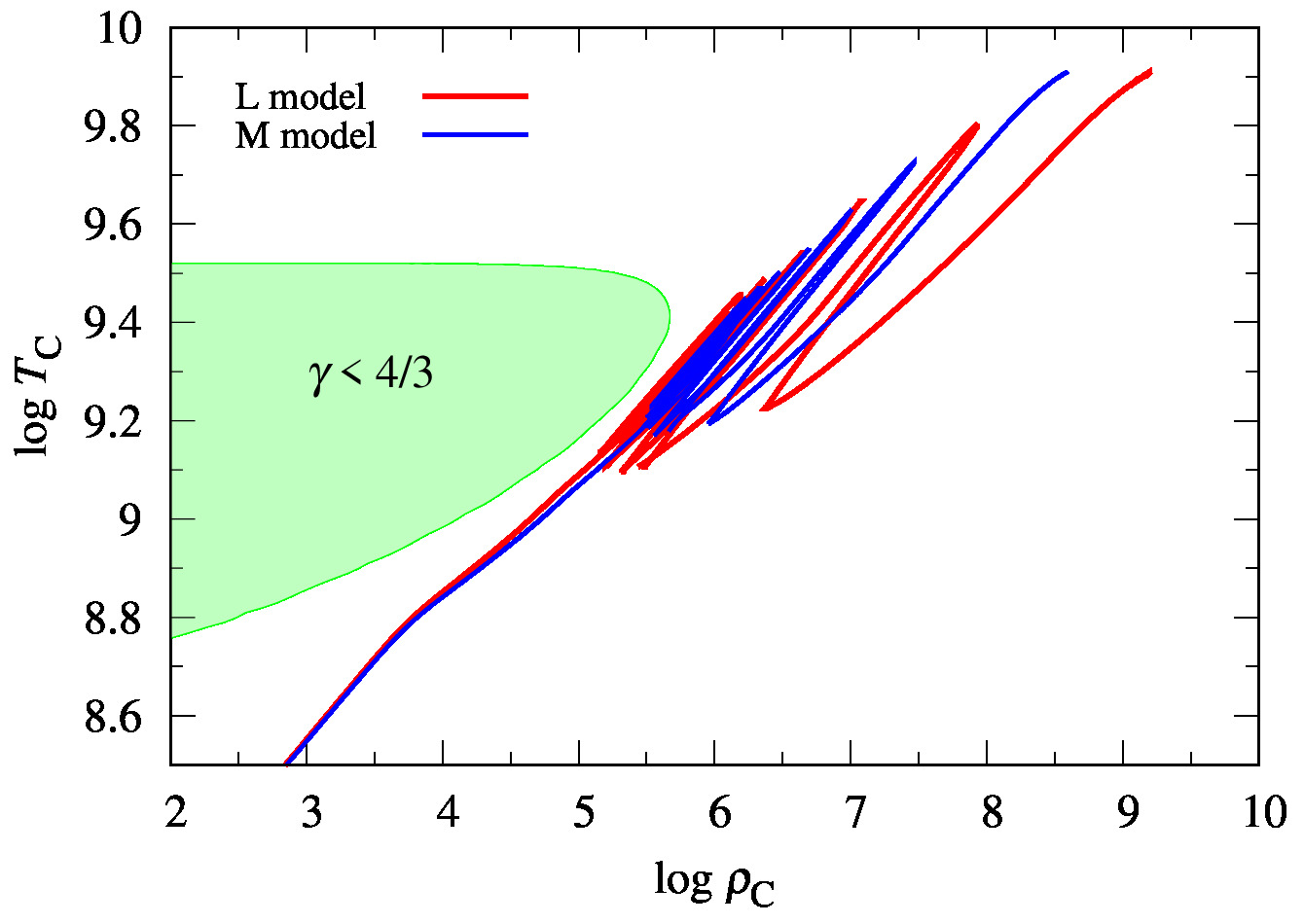}
\caption{The evolution of central temperature $T_{\rm c}$ and
central density $\rho_{\rm c}$ for the 100$M_\odot$ models around 
the PPI phase.
The red and blue curves indicate L and M models, respectively.
The green-shaded region is unstable against the electron-positron pair-instability where the adiabatic index $\gamma < 4/3$. 
\label{fig:2}}
\end{figure}

\begin{figure}[ht]
\includegraphics[width=8.5cm]{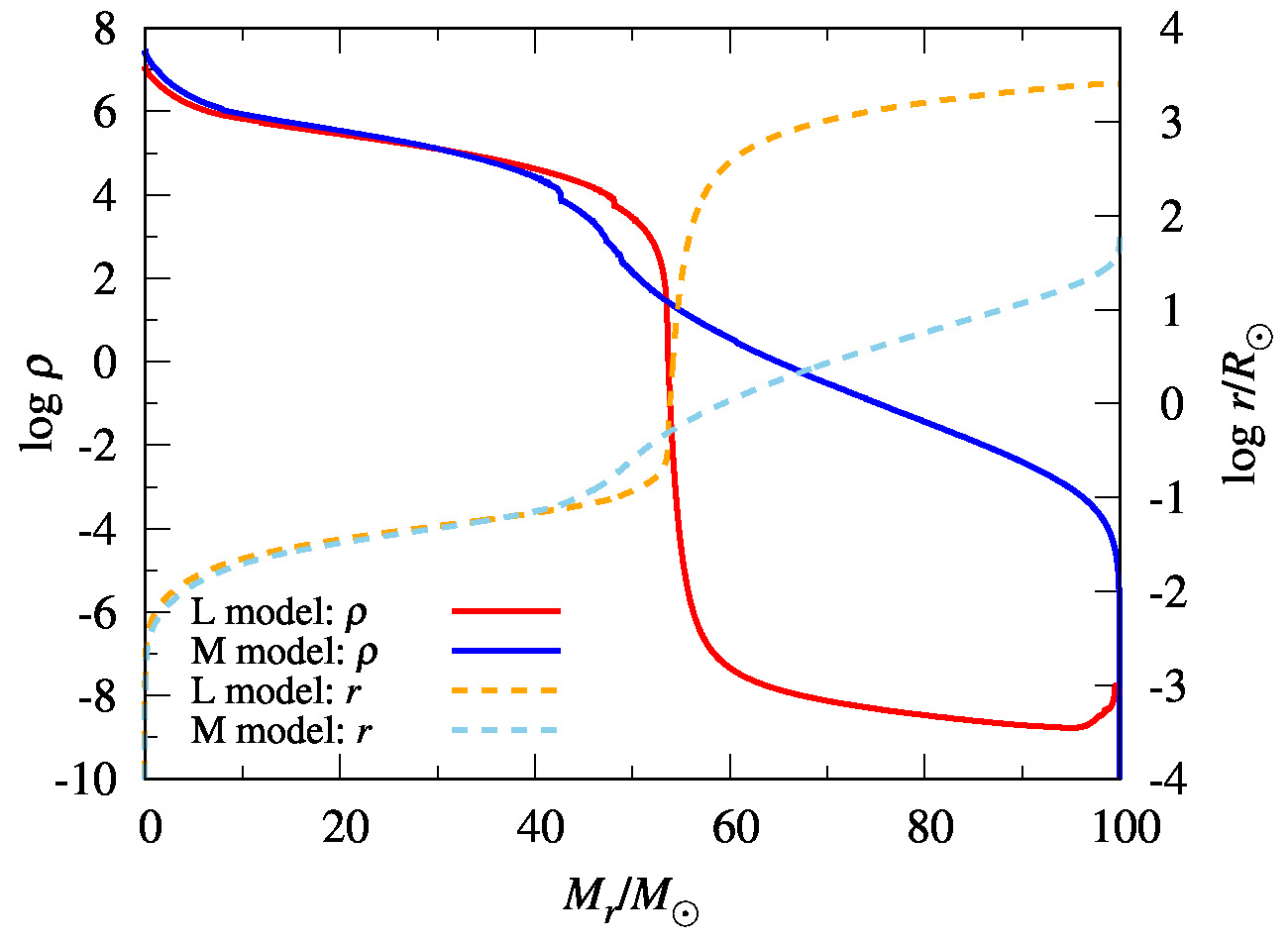}
\caption{The internal density and radius distribution
of the 100$M_\odot$ models
as a function of enclosed mass ($M_r$) at a time just before the first mass ejection.
Red and blue solid curves indicate the density ($\rho$) 
distributions for the L and M models.
Orange and skyblue dashed curves are the radius ($r$) 
distributions for the L and M models.
\label{fig:3}}
\end{figure}

\section{Implications \label{sec:implications}}


\subsection{L models \label{sec:Lmodels}}

 In the L models, the remnant mass, which we associate with the BH mass,
 ranges from 42.4 to 70 $M_\odot$ for the initial mass range
  of 70 to 135 $M_\odot$. The maximum BH mass is obtained for the
 lowest initial mass, 70 $M_\odot$, because only this case does not
 experience PPI mass ejection. Although only the last PPI pulse 
 ejects mass, almost all of the hydrogen envelope is removed
 by this pulse. As a result, the expected range for the BH mass 
 gap will be around 70 to 130 $M_\odot$ assuming the upper bound of the
 gap is determined by the most massive He core mass exploding as a pair-instability supernova \citep{2018ApJ...857..111T}. 
 The lower bound $\sim 70 M_\odot $ is a little
 larger than other arguments \citep[e.g.,][]{2020PhRvL.125j1102A}
 but is still marginaly small to explain 
 GW190521 ($M_1=85^{+21}_{-14} M_\odot $). 
 

\subsection{M models \label{sec:Mmodels}}

Although the parameter choice of the M models
is as reasonable as the L models, the remnant BH
masses are surprisingly different between the two models.
First, we find that $M_{\rm CO}$ of the M models
is smaller than that of the L models for the same initial mass.
This is because smaller $f_{\rm OV}$ means weaker convective 
mixing which leads to smaller He and CO cores. This smaller CO core
mass is one reason why the 80 $M_\odot$ model does not experience 
PPI mass ejection in the M model.

Though this effect is important, another effect is 
much more important in estimating the maximum BH mass 
below the mass gap.  
As described above, the L models have a much larger radius
than the M models during the PPI phase. Thus, the binding
energy of the hydrogen envelope of the M models 
($\sim 10^{51}$ erg) is typically
two orders of magnitudes larger than that of the
L models ($\sim 10^{49}$ erg). As shown in Table 1., the typical
ejecta energy is of the order of $\sim 10^{50}$ erg
and is too small to blow off the entire hydrogen envelope
of the M models. Therefore, the remnant mass of the M models 
is larger than that of the L models.
A zero-metallicity model calculated in \cite{2020arXiv200906585F} 
has a compact envelope similar to our models, but they concluded 
that the BH mass would be small because PPI occurred. 
We show, however, that the PPIs are not strong enough to eject mass.

In our results, the remnant mass monotonically increases from 70 to
 99 $M_\odot$ for $M_{\rm ini}$= 70 to 115 $M_\odot$.
 For $M_{\rm ini}$= 120 to 130 $M_\odot$ the remnant BH masses
 are smaller than the $M_{\rm ini}$= 115$M_\odot$ model because 
 PPI mass ejection occurs twice. The second mass
 ejection is much stronger than the first one. This is because the 
 binding energy of the hydrogen envelope is reduced by the first ejection, 
 and also because the peak temperature during the second PPI oscillation is higher, creating a stronger outgoing shock wave.
The maximum remnant mass is obtained
for $M_{\rm ini}$= 135$M_\odot$ for which the BH mass will be about 110$M_\odot$.
This is because this model collapses without the second mass ejection.
Therefore, the maximum BH mass below the mass gap could
be obtained for a model just below the pair-instability 
supernova region. We should stress that the M model's highest remnant mass corresponding to the highest initial mass
is quite different from the L model, in which the maximum BH mass
corresponds to the most massive model which does not experience
PPI mass ejection.

\subsection{GW190521 \label{sec:GW190521}}

After the discovery of GW190521, several
possibilities have been proposed to explain the large primary BH mass. 
Here we propose another simple solution by applying relatively weak convective overshooting mixing with a parameter of $f_{\rm OV}$=0.01.
In this M Model, a BH with $\sim 110 M_\odot$ may
be produced just below the BH mass gap.
 We note that our result is for
 single star evolution, and thus can be immediately applied to binary BH formation models without binary star interactions. 
 
 Recently \cite{2020arXiv201007616T} discussed that in their binary
 evolution models, the L models have difficulty explaining 
 the primary BH of GW190521. On the other
 hand the M models may explain it because they can
 avoid large mass loss during binary interactions due to their
 small radius. We note that they assume different relations
 between $M_{\rm ini}$ and $M_{\rm rem}$ from this work and they
 assume that no PPI occurs for $M_{\rm{He}} < 45 M_\odot$. Nevertheless
 their assumptions are not far from our M model, and become closer
 if the PPI pulses are weakened for some reason.

  There has been a suggestion that the PPI pulses are
  weakened if a smaller $^{12}$C$(\alpha, \gamma)^{16}$O
  rate \citep{2019ApJ...887...53F} is adopted. 
  The rate we use here is a standard
  value to explain the abundance of the universe. 
 We discuss elsewhere how much we can vary the rate to be
 consistent with the abundance observations.
 
 \subsection{PPI supernovae (SNe)\label{sec:Other}}

 In our results, no L models experience PPI mass ejection twice.
 These stars would shine as dark SNe IIp, since the
 ejecta energy is smaller than usual SNe.
 For M models, the stars would also shine as dark SNe II 
 during the first
 mass ejection. After the second mass ejection, $M= 120-130M_{\odot}$
stars would shine brighter since the ejecta energies are larger.  
These stars would shine one more time due to the collision between 
the first and the second ejecta. The light curves for these stages
will be investigated elsewhere.

\software{HOSHI \citep{2018ApJ...857..111T,Yoshida19}}

\acknowledgments
We thank A. Tanikawa and T. Kinugawa for useful discussions.
This research has been supported in part by Grants-in-Aid for
Scientific Research (17H01130, 17K05380, 19K03907, 20H05249)
from the Japan Society for the Promotion of Science.








\end{document}